\documentclass{PoS}
\usepackage{amsmath}
\usepackage{amssymb}
\usepackage{amsbsy}  
\usepackage{amsfonts}
\usepackage{graphicx}
\usepackage{subfigure}
\usepackage{latexsym}
\usepackage{mathrsfs}
\usepackage{dcolumn} 
\usepackage{wasysym}
\usepackage{url}

\title{Spanning of Topological sectors, charge and
susceptibility with naive Wilson fermions}

\ShortTitle{}

\author{Abhishek Chowdhury, Asit K. De\thanks{Speaker}, Sangita De Sarkar,
 A. Harindranath, Santanu Mondal, Anwesa Sarkar\\
        Theory Division, Saha Institute of Nuclear Physics \\
 1/AF Bidhan Nagar, Kolkata 700064, India}

\author{Jyotirmoy Maiti\\
        Department of Physics, Barasat Government College,\\
10 KNC Road, Barasat, Kolkata 700124, India}

\abstract{We study the topological charge and the topological susceptibility 
in lattice QCD with two 
degenerate flavors of naive Wilson fermions at two values of lattice spacings
and different volumes, for a range of quark masses. Configurations are 
generated with DDHMC/HMC algorithms and smoothened with HYP smearing. 
We present integrated autocorrelation time for
both topological charge and topological susceptibility at the two lattice spacing
values studied. The spanning of different topological sectors as a 
function of the hopping parameter $\kappa$ is presented. The expected chiral behaviour of the 
topological susceptibility (including finite volume dependence) is observed.}

\FullConference{ The XXIX International Symposium on Lattice Field Theory - 
Lattice 2011\\
July 10-16, 2011\\
Squaw Valley, Lake Tahoe, California}

\begin{document}

\section{Introduction}
Earlier attempt \cite{bali} in lattice QCD to verify the suppression of topological 
susceptibility with decreasing quark 
mass ($m_q$) with unimproved Wilson fermions and HMC algorithm 
was unable to unabiguously confirm the suppression. 
Since topological 
susceptibility is a measure of the spanning of different topological sectors 
of QCD vacuum, the inability to reproduce the predicted suppression may raise 
concerns about the simulation algorithm and the particular fermion formulation 
to span the configuration space correctly and/or efficiently.  
In this work, we perform a systematic study using unimproved Wilson fermions
and demonstrate  the  
suppression of topological susceptibility with decreasing quark mass.
This work is  part of an ongoing program \cite{anomaly1, anomaly2}
to study the chiral properties of 
Wilson lattice QCD. For details on measurements of topological charge and 
susceptibility, 
and numerical values of measured quark masses, pion masses and 
topological susceptibilities see Ref. \cite{topo1}.
Our results of the topological susceptibility favourably
compare in Ref. \cite{topo1} with that from a recent mixed action calculation employing clover 
fermions and overlap fermions for the sea and the valence sectors respectively 
\cite{bernardoni}. 
The detailed account of low lying spectroscopy will appear
separately.

\section{Measurements}

\begin{figure}
\subfigure{   
\includegraphics[width=2.5in]{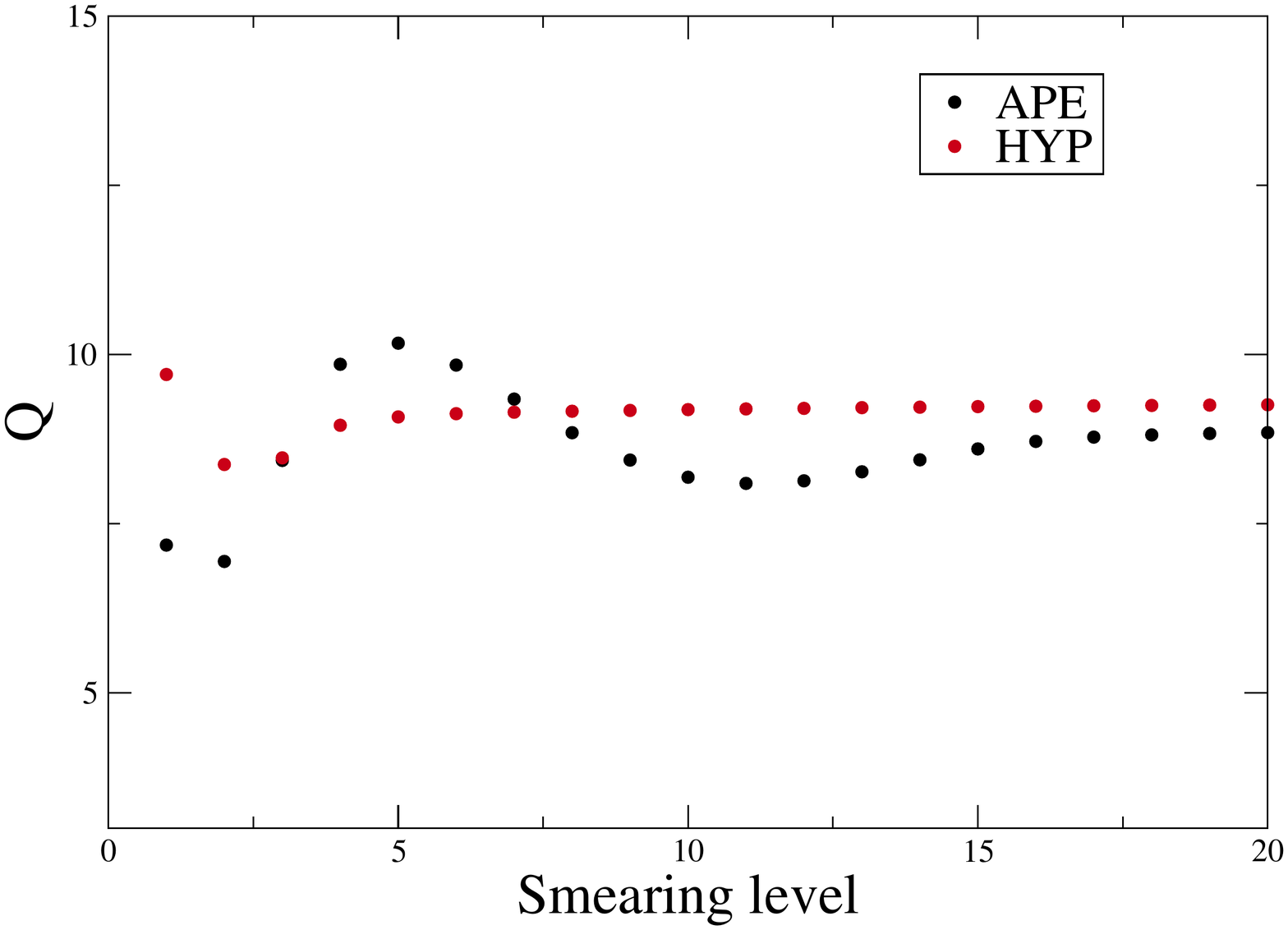}
}
\subfigure{
\includegraphics[width=2.5in]{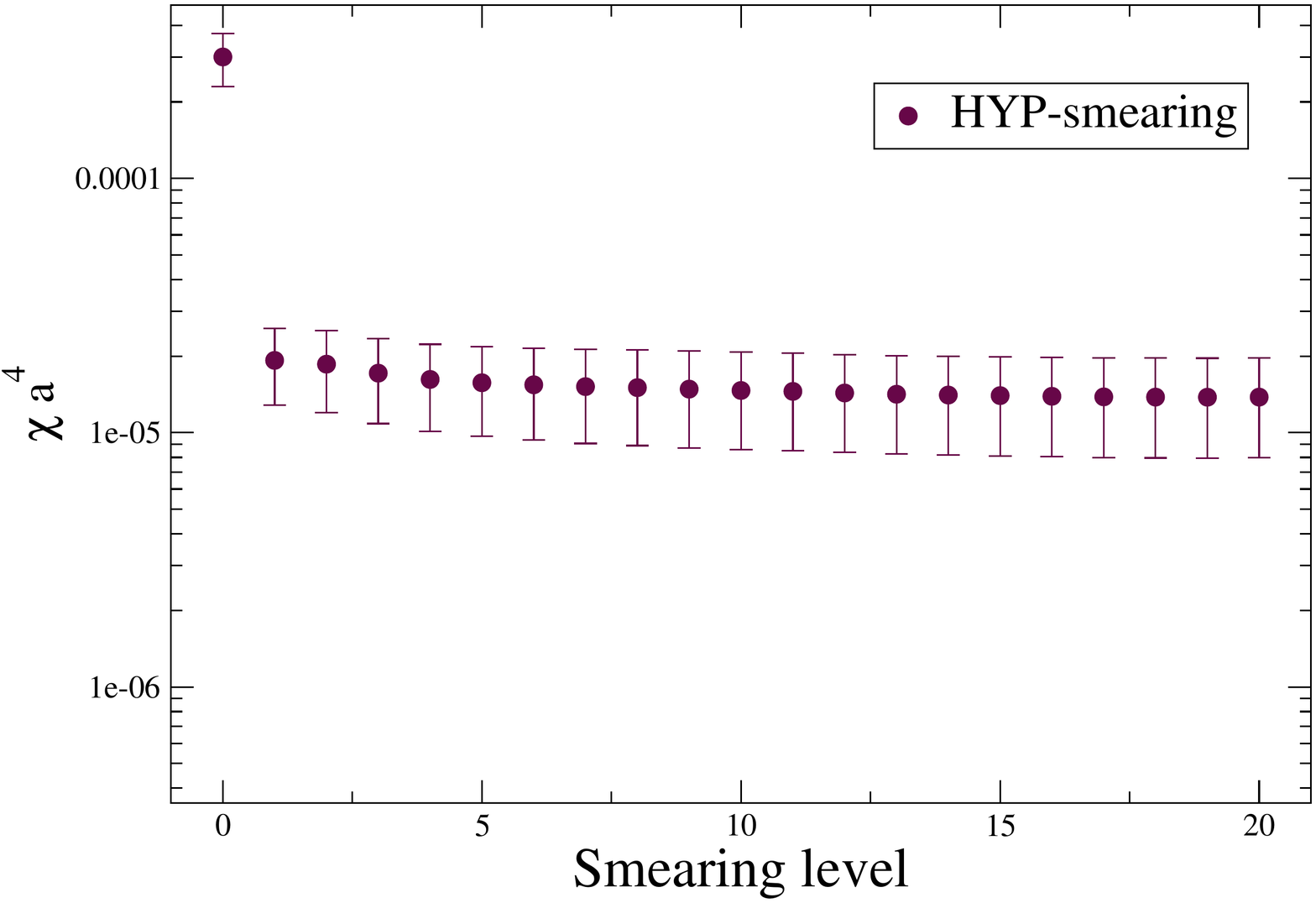}
}
\caption{(a) Topological charge versus smearing levels for APE  and 
HYP smearing at $\kappa = 0.15825$, $\beta = 5.6$ and volume $24^3 \times 48$.
(b) Topological susceptibility versus HYP smearing levels at 
$\kappa = 0.15825$, $\beta = 5.6$ and volume $24^3 \times 48$. }
\label{fig1}
\end{figure}

\begin{figure}
\begin{center}
\subfigure{   
\includegraphics[width=2.5in]{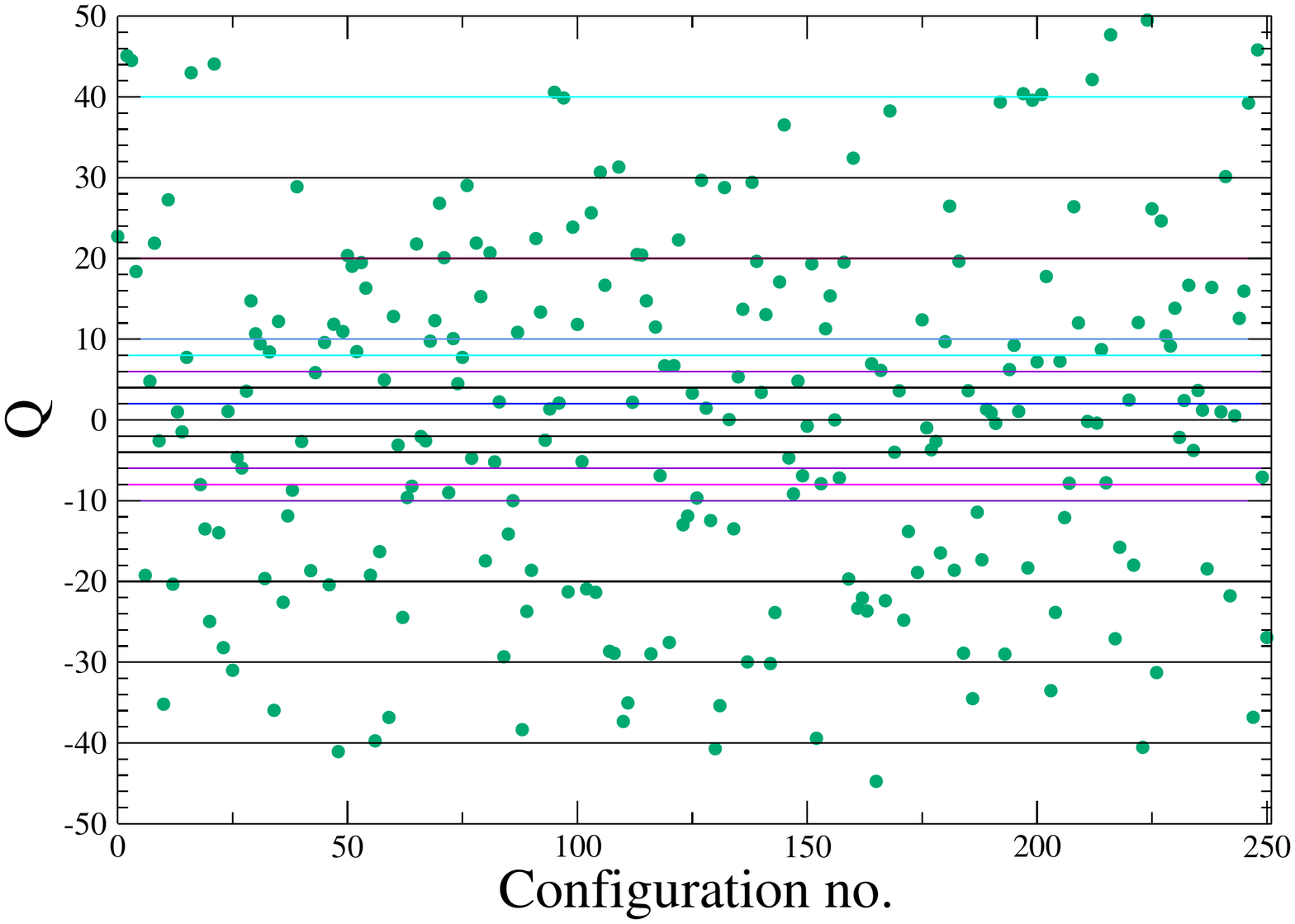}
}
\subfigure{
\includegraphics[width=2.5in]{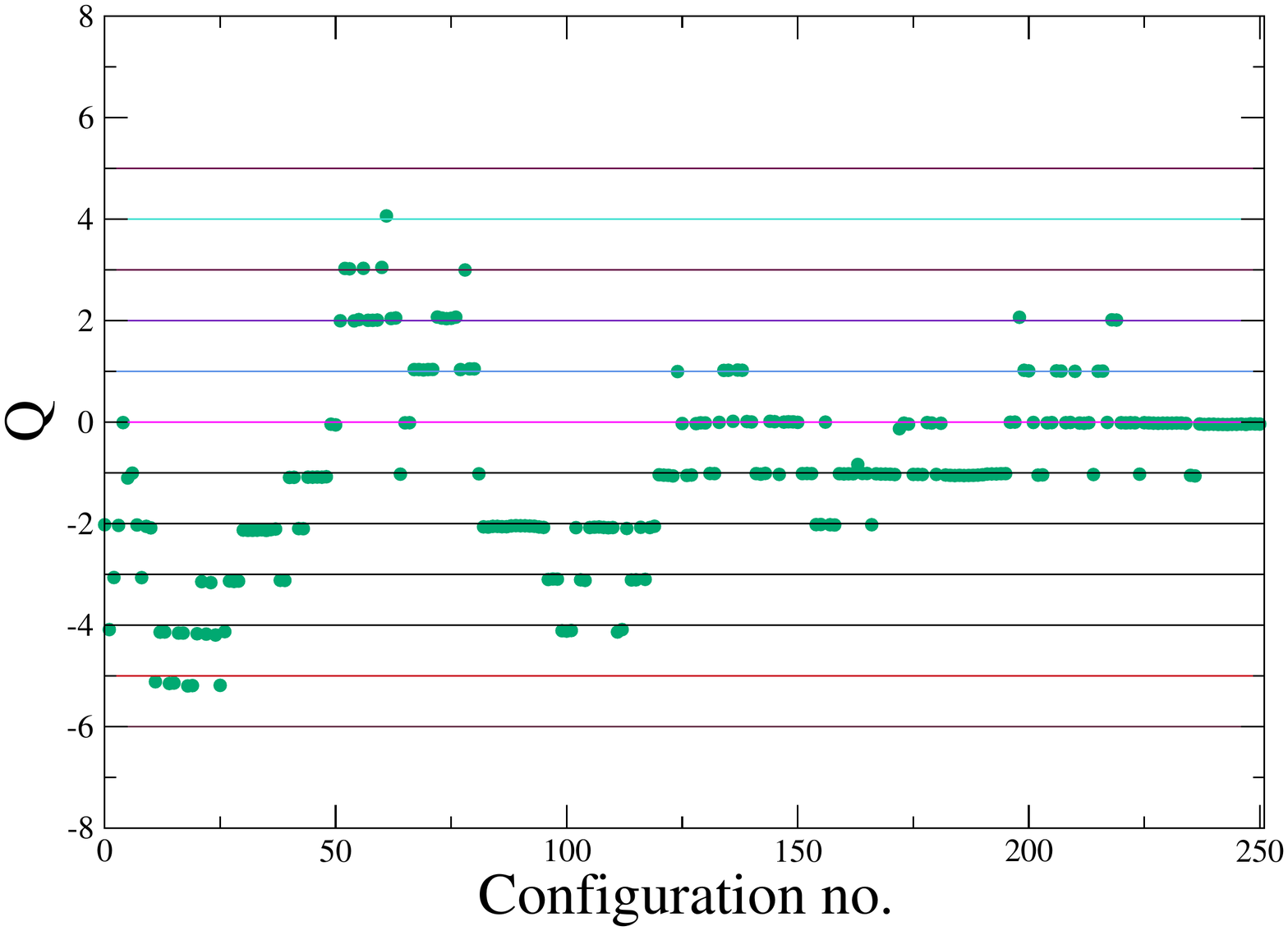}
}
\caption{Topological charge for different configurations at volume 
 $32^3 \times 64$, 
$\beta = 5.8$ and $\kappa = 0.15475$ with (a) no smearing (b) $20$ 
levels of HYP smearing.}
\label{fig2}
\end{center}
\end{figure}

\begin{figure}
\begin{center}
\subfigure{
\includegraphics[width=2.5in]{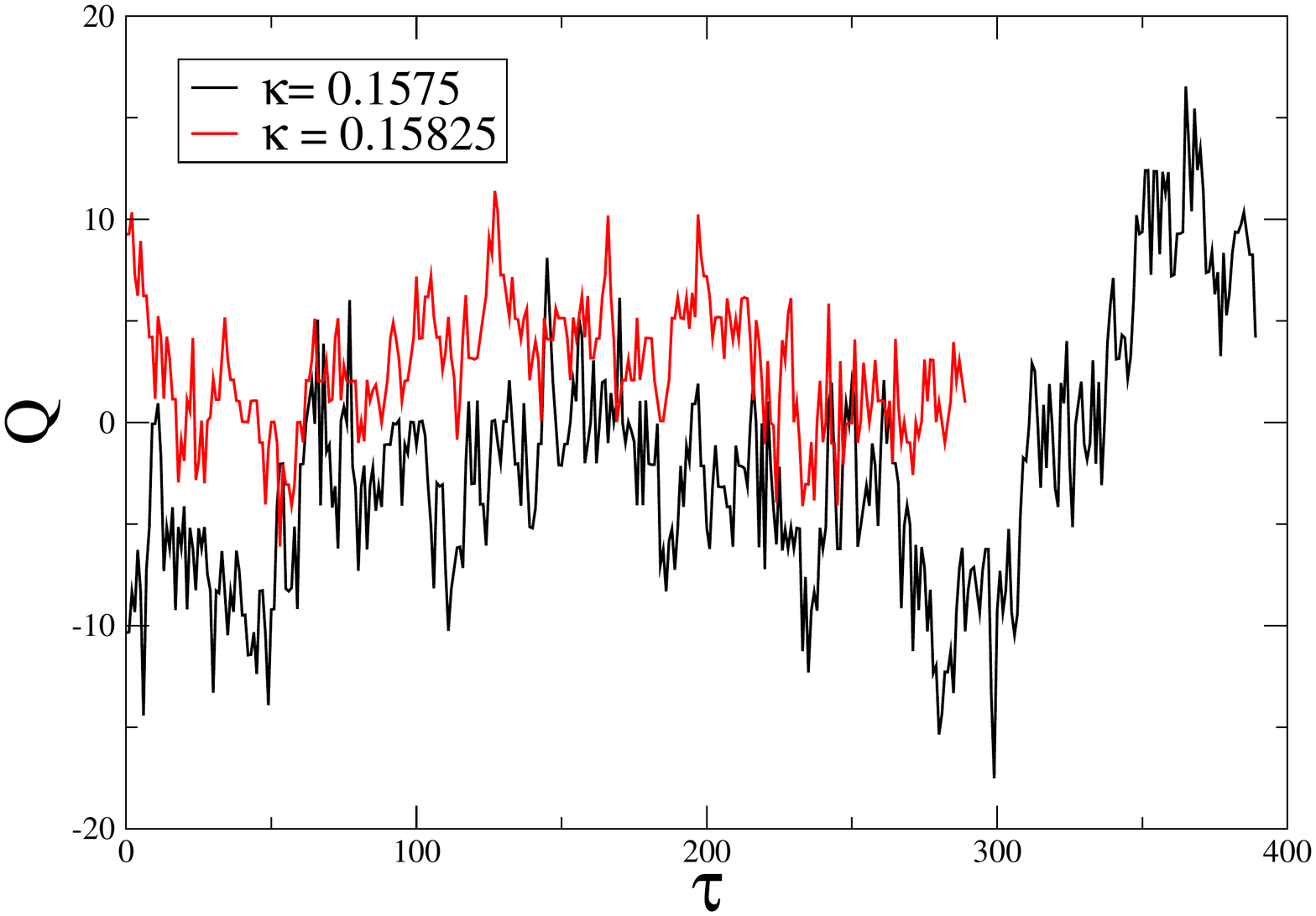}
}
\subfigure{
\includegraphics[width=2.5in]{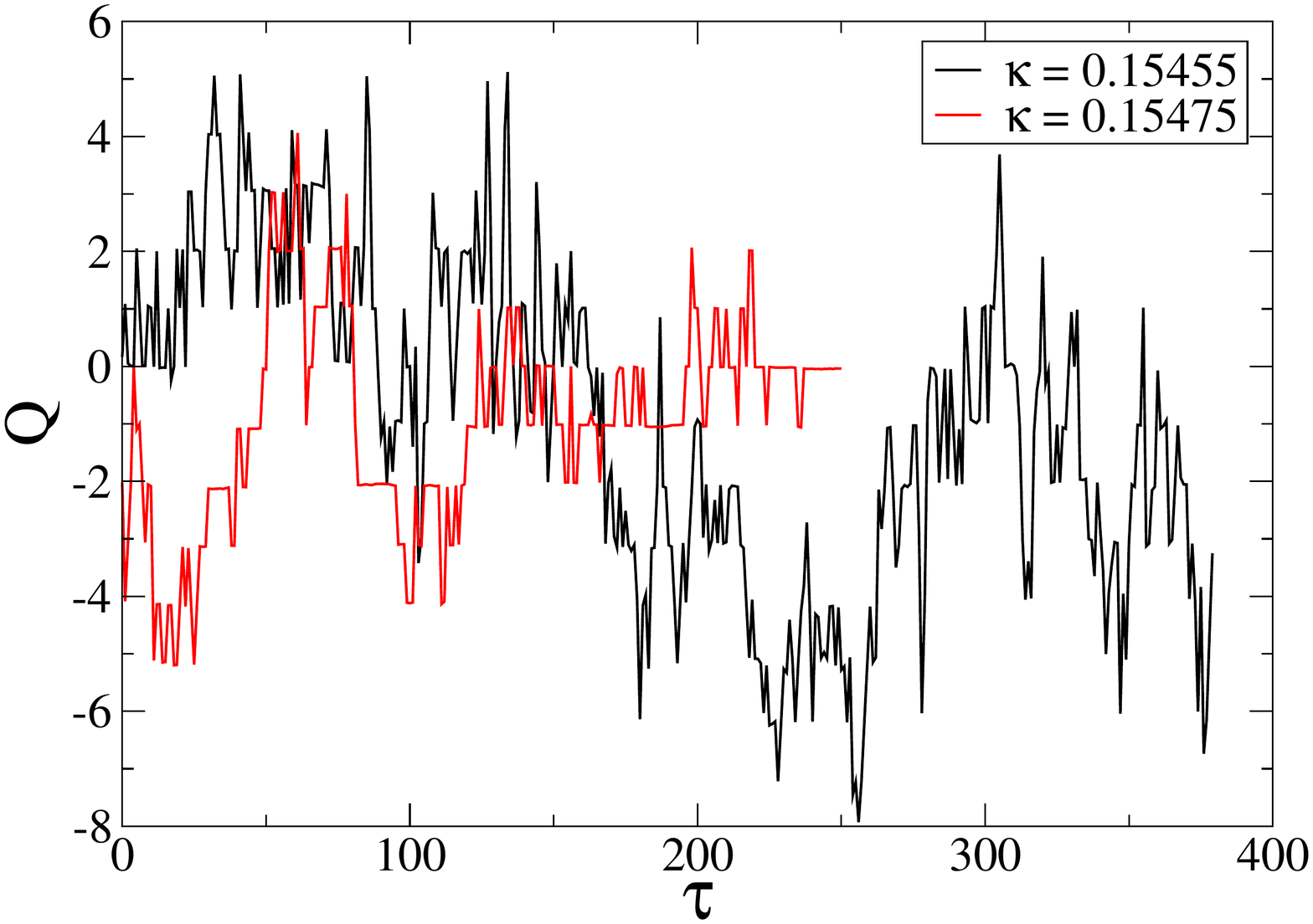}
}
\caption{The Monte Carlo trajectory history for topological charge with 
unimproved Wilson fermion and gauge action for (a) $\beta = 5.6$, volume $=24^3\times 48$ 
with a gap of 24 trajectories between two consecutive measurements  
and 
(b) $\beta = 5.8$, volume $=32^3\times 64$ 
with a gap of 32 trajectories between two consecutive measurements.}
\label{fig3}
\end{center}
\end{figure}

\begin{figure}
\begin{minipage}{2\linewidth}
\includegraphics[width=2.1in]{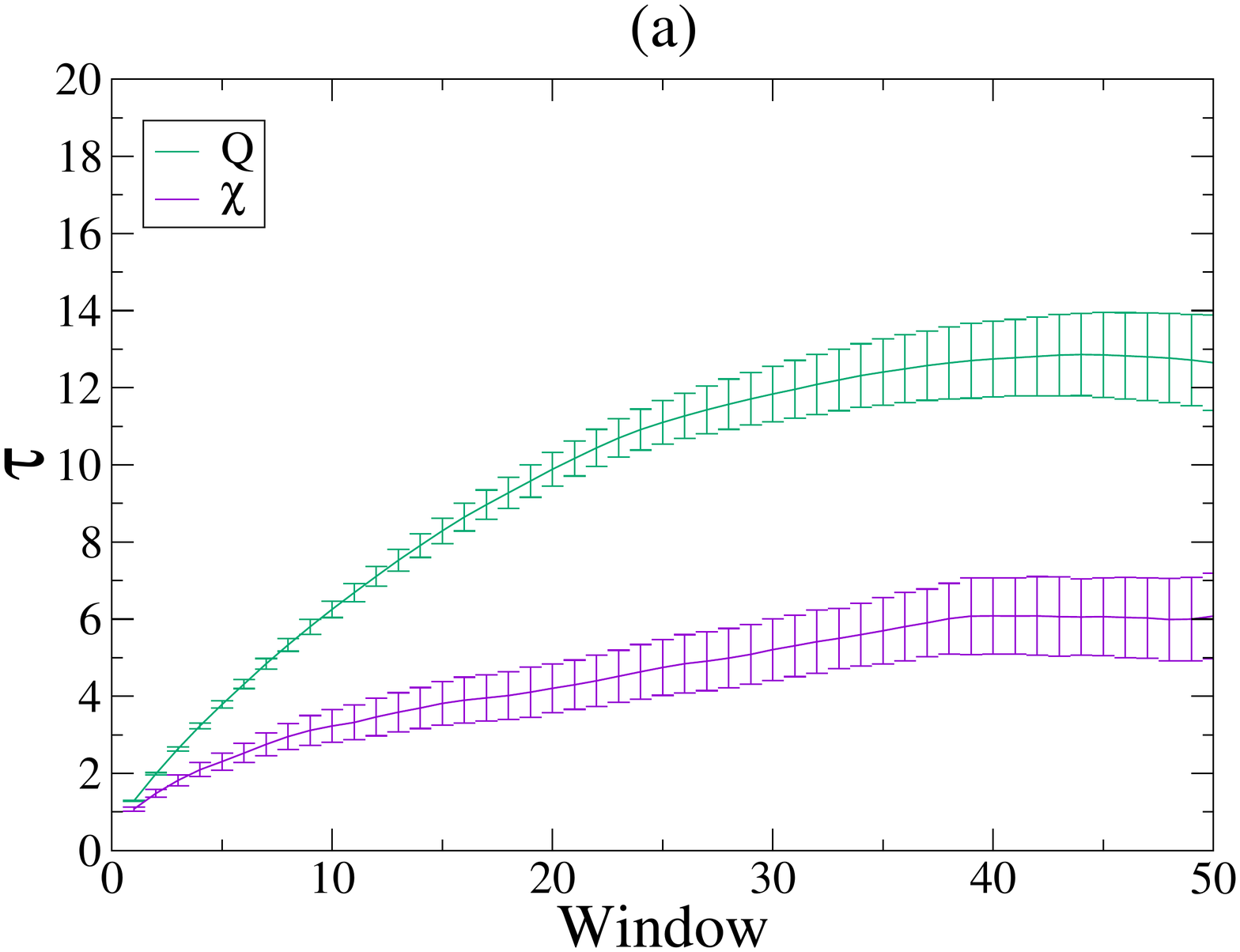}
\includegraphics[width=2.1in]{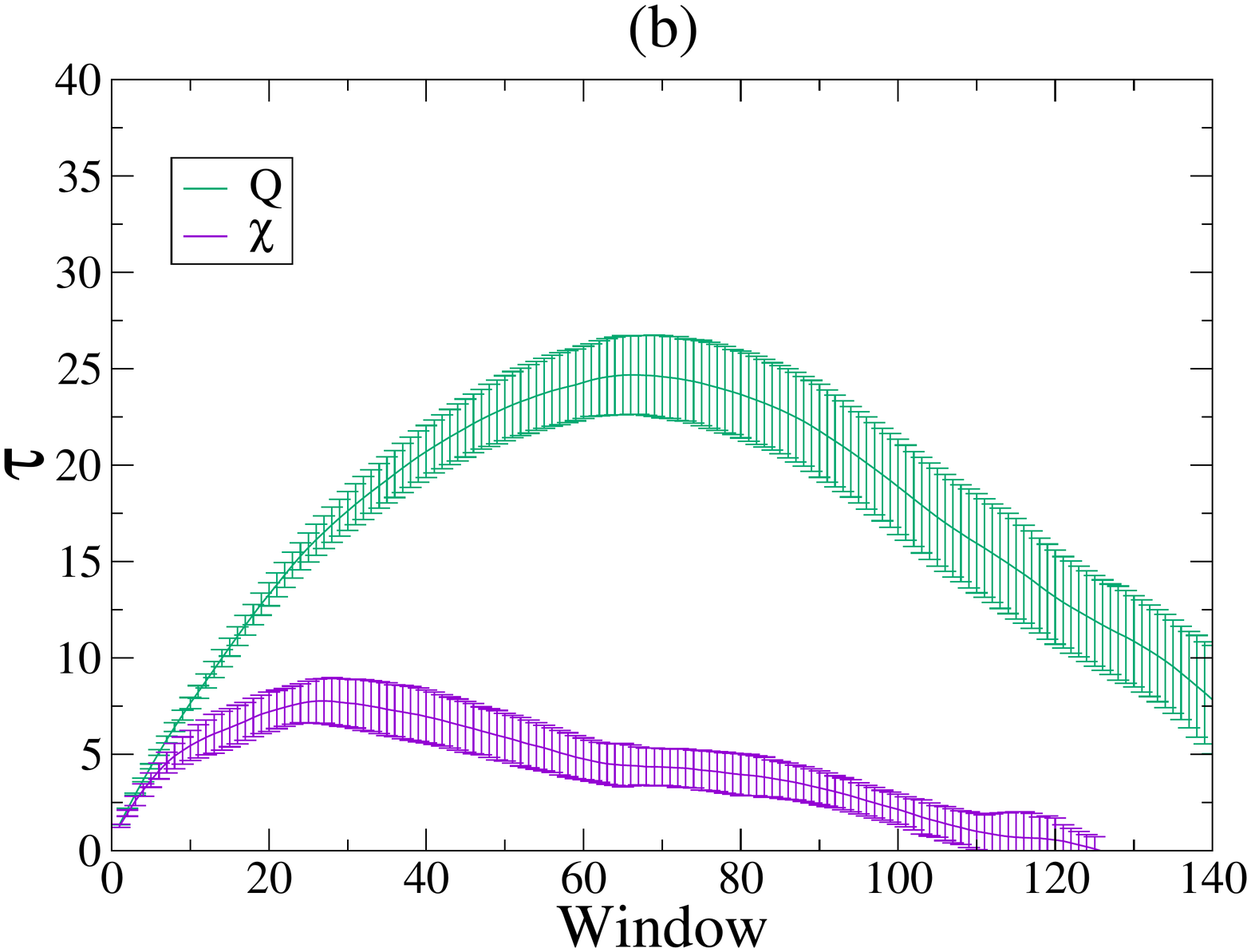}
\includegraphics[width=2.1in]{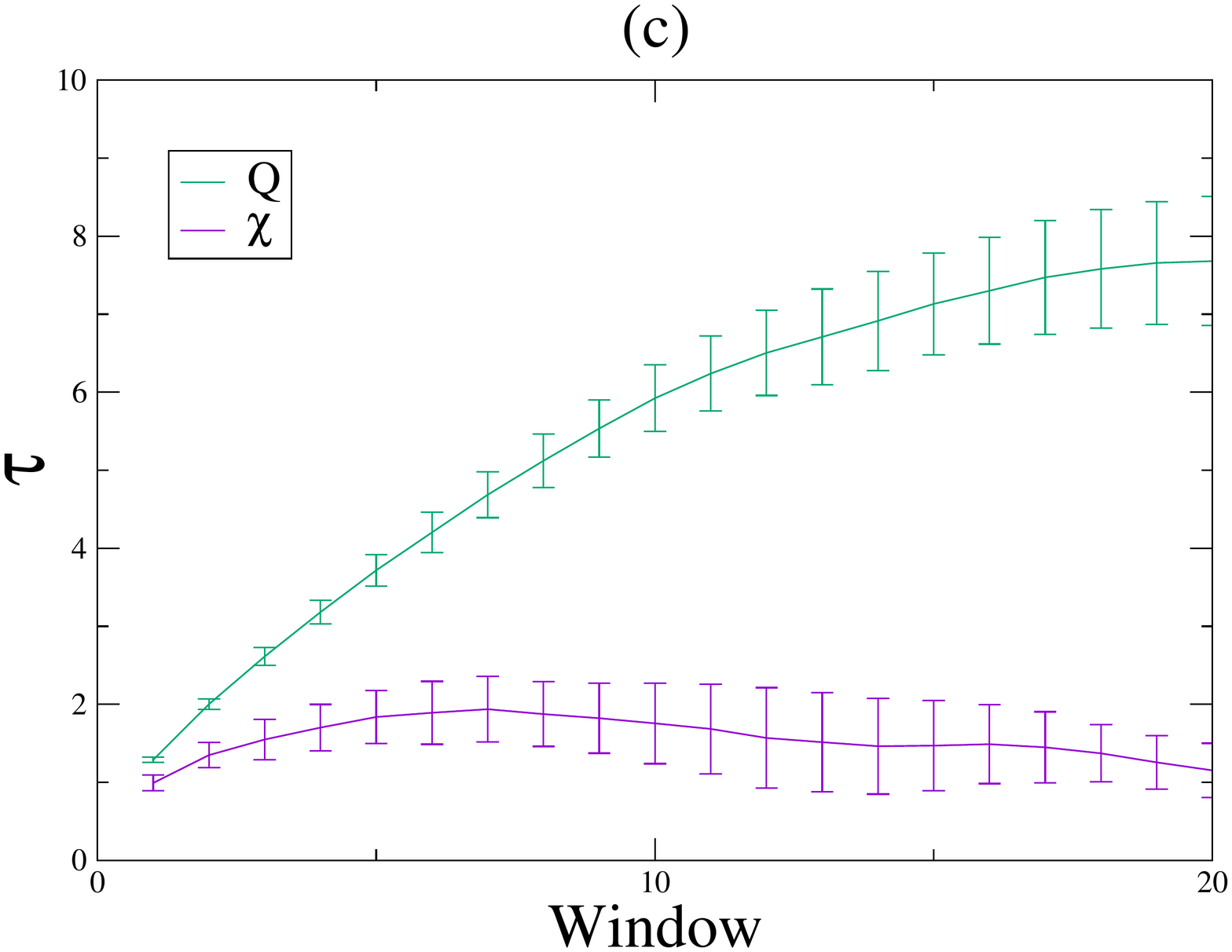}
\end{minipage}
\begin{minipage}{2\linewidth}
\includegraphics[width=2.1in]{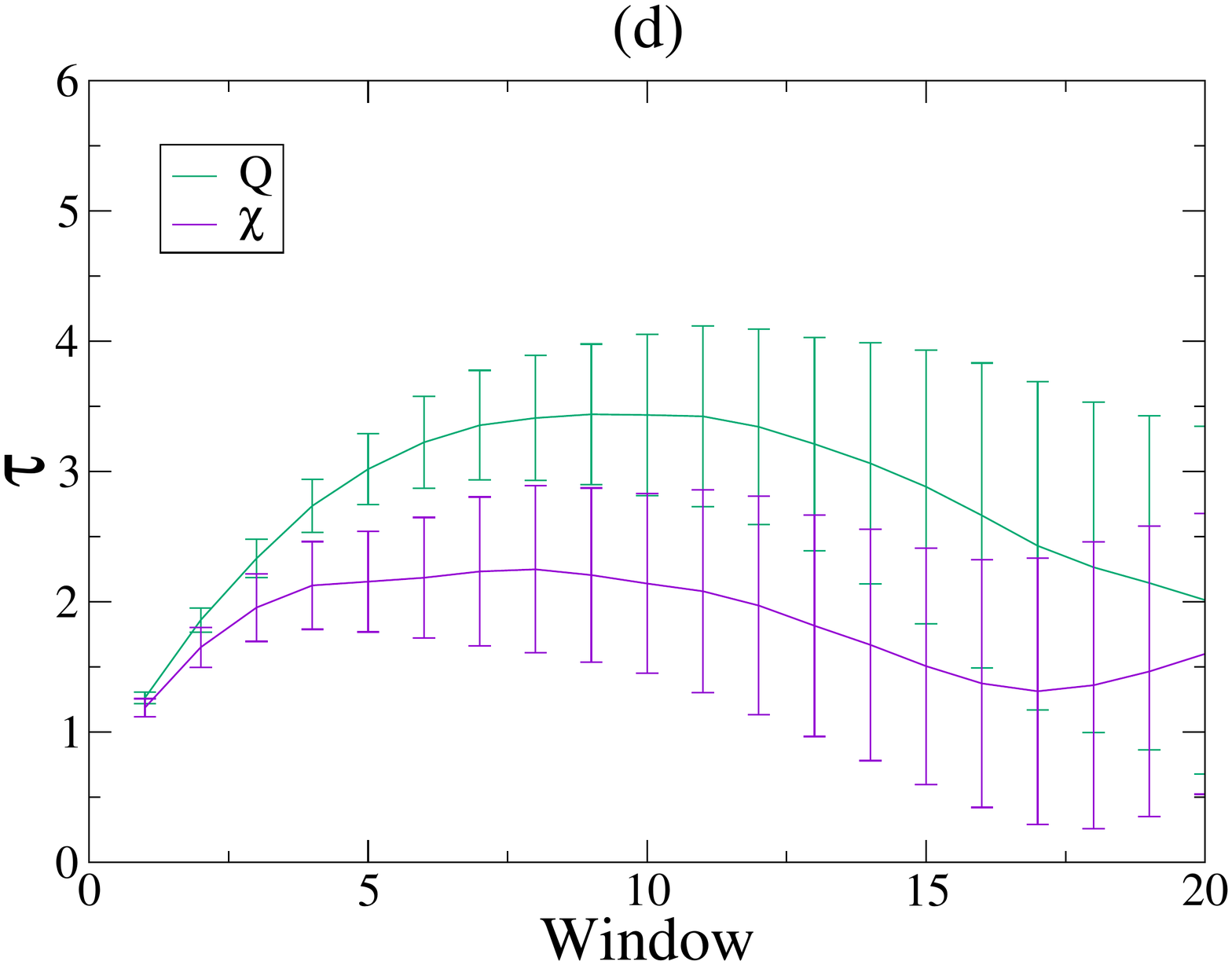}
\includegraphics[width=2.1in]{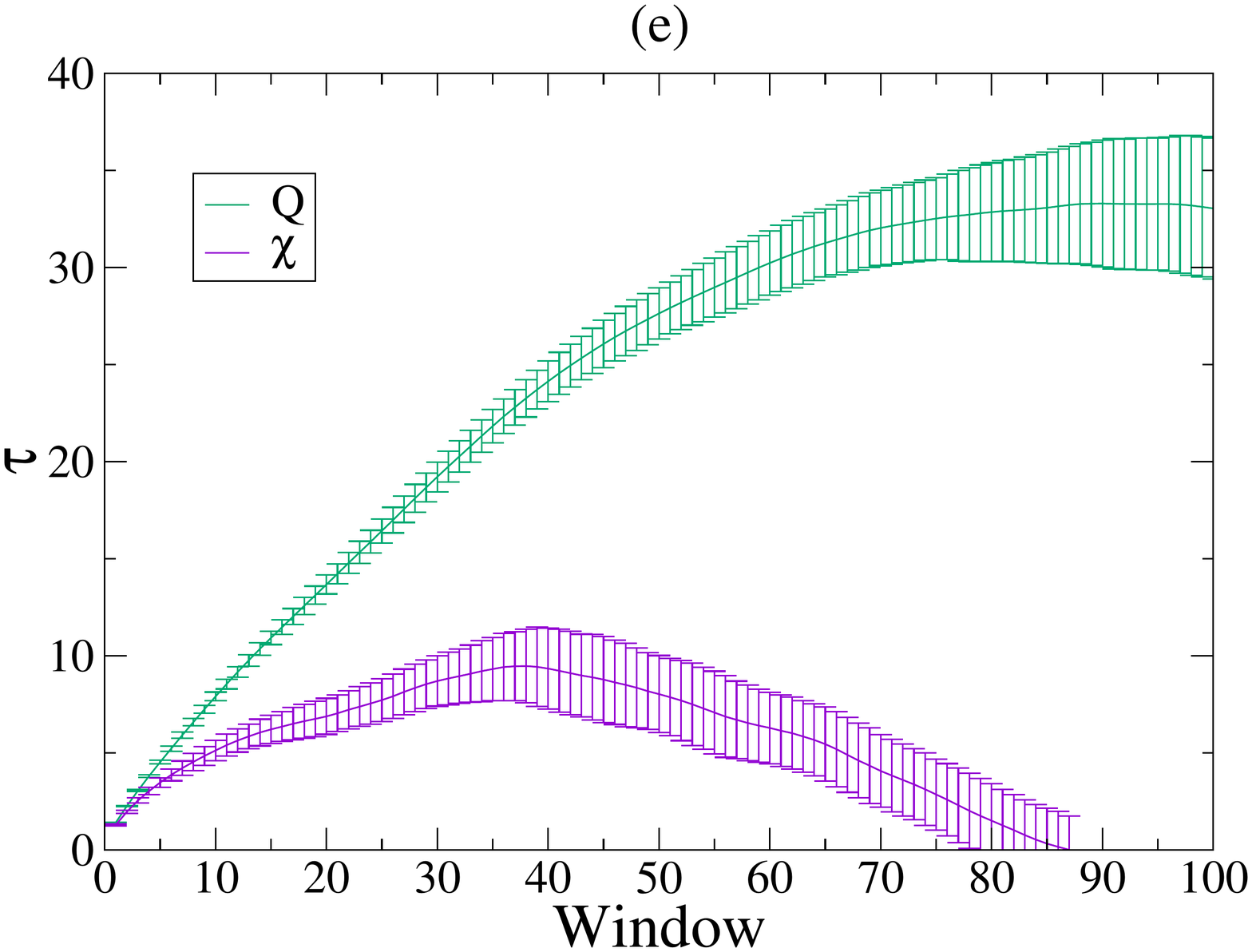}
\includegraphics[width=2.1in]{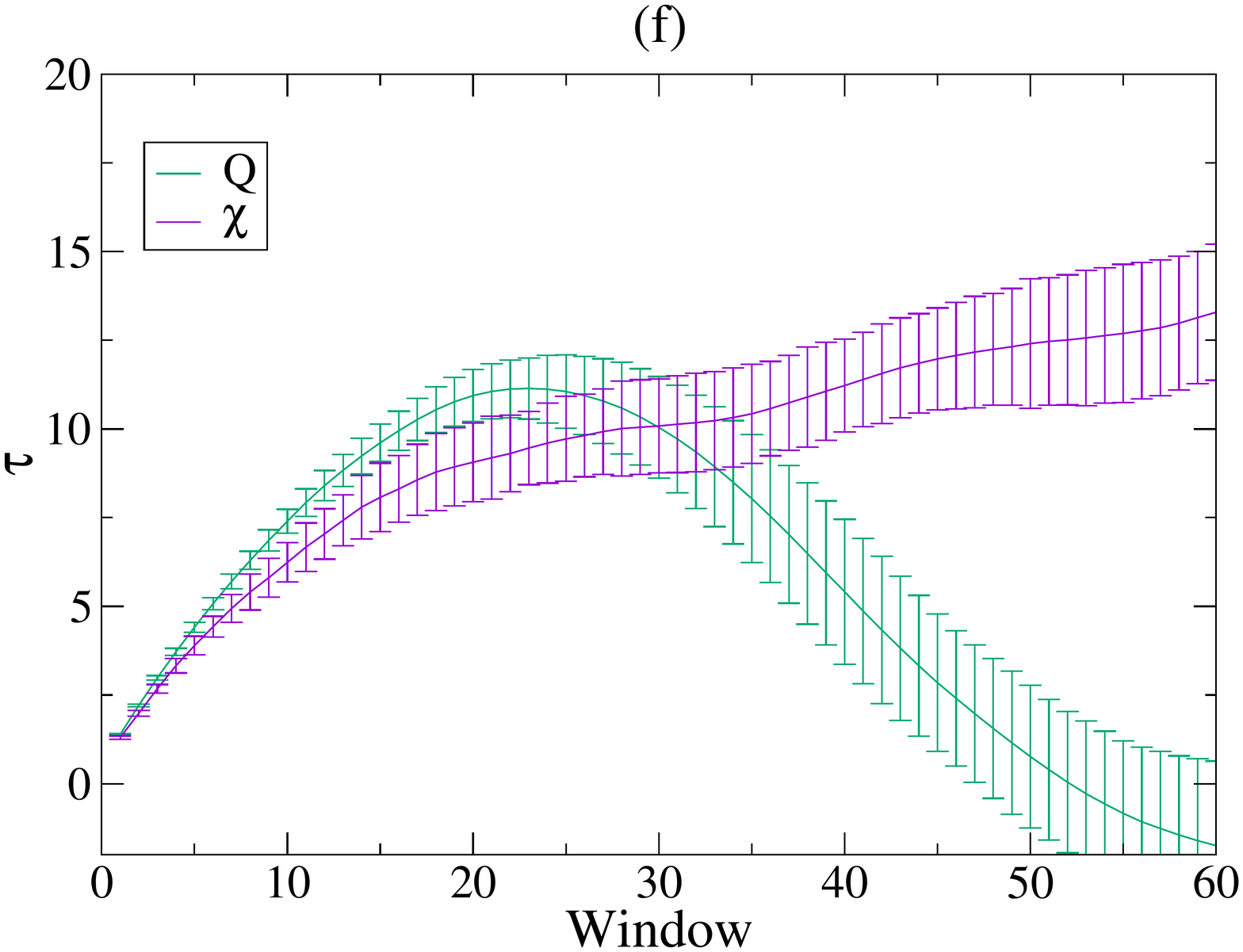}
\end{minipage}
\caption{The integrated autocorrelation lengths 
for topological charge and susceptibility at 
(a) $\beta = 5.6$, $\kappa = 0.1575$, volume $=16^3\times 32$
(b) $\beta = 5.6$, $\kappa = 0.15775$, volume $=24^3\times 48$
(c) $\beta = 5.6$, $\kappa = 0.15775$, volume $=32^3\times 64$
(d) $\beta = 5.6$, $\kappa = 0.15815$, volume $=32^3\times 64$
(e) $\beta = 5.8$, $\kappa = 0.15455$, volume $=32^3\times 64$
(f) $\beta = 5.8$, $\kappa = 0.15475$, volume $=32^3\times 64$.
The absolute scales for both the axes are obtained by multiplying by 25 in (a), by 24 in (b) and by 32 in the rest.}
\label{fig4}
\end{figure}

We have generated ensembles of gauge configurations by means of HMC
\cite{hmc} and DDHMC \cite{ddhmc}
algorithms using unimproved Wilson fermion and gauge actions with 
$n_f=2$ mass degenerate quark flavours at two values of the gauge coupling ($\beta=$5.6, 5.8) . At $\beta=5.6$ the lattice volumes are 
$16^3 \times 32$, $24^3 \times 48$ and $32^3 \times 64$ and the renormalized 
quark mass varies from $15$ to $100$ MeV ($\overline{\rm MS}$ scheme 
 at $2$ GeV). Quark masses are determined using axial Ward identity. At $\beta = 5.8$ the lattice 
volume is $32^3 \times 64$ and the renormalized quark mass ranges 
from $20$ to $90$ MeV. The lattice spacings determined using Sommer parameter
at $\beta =5.6$ and $5.8$ are $0.077$ and $0.061$ fm respectively. 
All the configurations for lattice volumes $24^3 \times 48$ and $32^3 \times 64$  
are generated using DDHMC
algorithm. The $16^3 \times 32$ configurations were generated using the HMC algorithm except for 
$\kappa = 0.15775$ where DDHMC was also used with different block sizes.
The number of thermalized configurations ranges from $2000$ to $12000$ 
and the number of measured configurations ranges from $70$ to $500$.
For topological charge density, we use the lattice approximation developed for 
$SU(2)$ by DeGrand, Hasenfratz and Kovacs \cite{degrand}, modified for
$SU(3)$ by Hasenfratz and Nieter \cite{hasenfratz1} and implemented in
the MILC code \cite{milc}.
To suppress the ultraviolet lattice artifacts, smearing of link fields 
is employed. The comparison of the effect of APE \cite{ape} and HYP smearing
\cite{hasenfratz2} on topological charge is presented in Fig. \ref{fig1}.
It is clear that HYP smearing is more effective than APE. 
We used  $20$ HYP smearing steps
with optimized smearing coefficients $\alpha =0.75$,
$\alpha_2=0.6$ and $\alpha_3=0.3$ \cite{hasenfratz2}. In Fig. \ref{fig2} we show that the topological 
charges are not close to integers before smearing but they are very close to integers after $20$ steps 
of HYP smearing.
Fig. \ref{fig3} shows Monte Carlo trajectory history of topological charge at $\kappa = 0.1575, 0.15825$,
 $\beta = 5.6$, volume $=24^3\times 48$ and at $\kappa = 0.15455, 0.15475$, $\beta = 5.8$,
volume $=32^3\times 64$. There is some sign of trapping
only at $\kappa = 0.15475$ at $\beta = 5.8$.
\begin{table}
\begin{center}
\begin{tabular}{|l|l|l|l|l|l|l|l|l|}
\hline 
\hline 
$\beta=5.6$ & \multicolumn{7}{c|}{} \\
\hline
 &$lattice$& $\kappa$& $block$& {$N_{trj}$} & {$N_{cfg}$} &{$\tau$}&
{$\tau_{int}^{\chi}$}&{$\tau_{int}^{Q}$} \\
\hline
&{   $16^3\times32$}&{$0.156$}&{HMC} &{5000} &{200}&{0.5} &{12(5)}&{41(6)} \\
&{   $~~~~~,,$}&{$0.157$}&{HMC}& {5000} &{200}&{0.5} &{15(7)}&{46(10)} \\
&{   $~~~~~,,$}&{$0.1575$}&{HMC}& {5000} &{1000}&{0.5} &{30(5)}&{64(5)} \\
&{$~~~~~,,$}&{$0.15775$}&{HMC}& {5000} &{1000}&{0.5} &{16(2)}&{58(6)} \\
&{$~~~~~,,$}&{$0.15775$}&{$8^4$} &{16320} &{510}&{0.5} &{186(32)}&{208(32)} \\
&{$~~~~~,,$}&{$0.15775$}&{$8^3\times 16$} &{9408} &{294}&{0.5} &{70(19)}&{213(49)} \\
&{   $~~~~~,,$}&{$0.158$}&{HMC} &{5000} &{200}&{0.5} &{18(4)}&{32(8)} \\
&{   $24^3\times48$}&{$0.1575$}&{$6^3\times 8$} &{9360} &{390}&{0.5} &{182(18)}&{468(52)} \\
&{   $~~~~~,,$}&{$0.15775$}&{$6^3\times 8$} &{10560} &{440}&{0.5} &{185(26)}&{590(48)} \\   
&{   $~~~~~,,$}&{$0.158$}&{$6^3\times 8$}& {7200} &{300}&{0.5} &{132(22)}&{290(30)} \\      
&{   $~~~~~,,$}&{$0.158125$}&{$6^3\times 8$} &{11760} &{490}&{0.5} &{149(22)}&{242(24)} \\  
&{   $~~~~~,,$}&{$0.15825$}&{$6^3\times 8$}& {6960} &{290}&{0.5} &{98(12)}&{259(34)} \\     
&{   $32^3\times64$}&{$0.15775$}&{$8^3\times 16$} &{6112} &{191}&{0.5} &{61(13)}&{243(25)} \\
&{$~~~~~,,$}&{$0.158$}&{$8^3\times 16$} &{4992} &{156}&{0.5} &{74(16)}&{211(30)} \\
&{$~~~~~,,$}&{$0.15815$}&{$8^3\times 16$} &{5024} &{157}&{0.5} &{70(13)}&{109(14)} \\
&{$~~~~~,,$}&{$0.1583$}&{$8^3\times 16$} &{2240} &{70}&{0.25} &{29(11)}&{36(8)} \\   
\hline \hline
\hline
$\beta=5.8$ & \multicolumn{7}{c|}{} \\
\hline
 &$lattice$& $\kappa$& $block$&  {$N_{trj}$} & {$N_{cfg}$} &{$\tau$}&
{$\tau_{int}^{\chi}$}&{$\tau_{int}^{Q}$} \\
\hline
&{$32^3\times64$}&{$0.1543$}& {$8^3\times 16$}&{9600} &{300}&{0.5} &{672(73)}&{2560(64)} \\
&{$~~~~~,,$}&{$0.15455$}& {$8^3\times 16$} &{12160} &{380}&{0.5} &{288(56)}&{1056(84)} \\   
&{$~~~~~,,$}&{$0.15475$}& {$8^3\times 16$} &{8032} &{251}&{0.5, 0.25}  &{310(45)}&{352(29)} \\
\hline \hline
\end{tabular}
\end{center} 
\caption{Lattice parameters and simulation statistics, $\tau$ is molecular dynamics trajectory length.}
\label{lattice}
\end{table}
In table \ref{lattice} we present, the simulation parameters and the measured
integrated autocorrelation times for topological susceptibility 
($\tau_{int}^{\chi}$) and topological charge ($\tau_{int}^Q$) (Two points shown in the presentation were preliminary and are currently going through more configuration generation. On the other hand, a few points on the lowest volume have been added, especially with different block size of DDHMC). There is some indication that $\tau_{int}^{\chi}$ and $\tau_{int}^Q$
decrease with increasing $\kappa$. The $\tau_{int}^{\chi}$ at volume $=16^3\times 32$, $\beta = 5.6$, $\kappa = 0.15775$ decreases 
with increasing block size. The $\tau_{int}^{\chi}$ at volume $=32^3\times 64$, $\beta = 5.6$ is lower than $24^3 \times 48$
and $16^3 \times 32$ DDHMC runs because of the higher block size which results in higher active to total link ratio \cite{sommer}.  
 In Fig. \ref{fig4} we present a few plots for $\tau_{int}^{\chi}$
and $\tau_{int}^Q$ determinations.

\begin{figure}
\subfigure{
\includegraphics[width=2.5in]{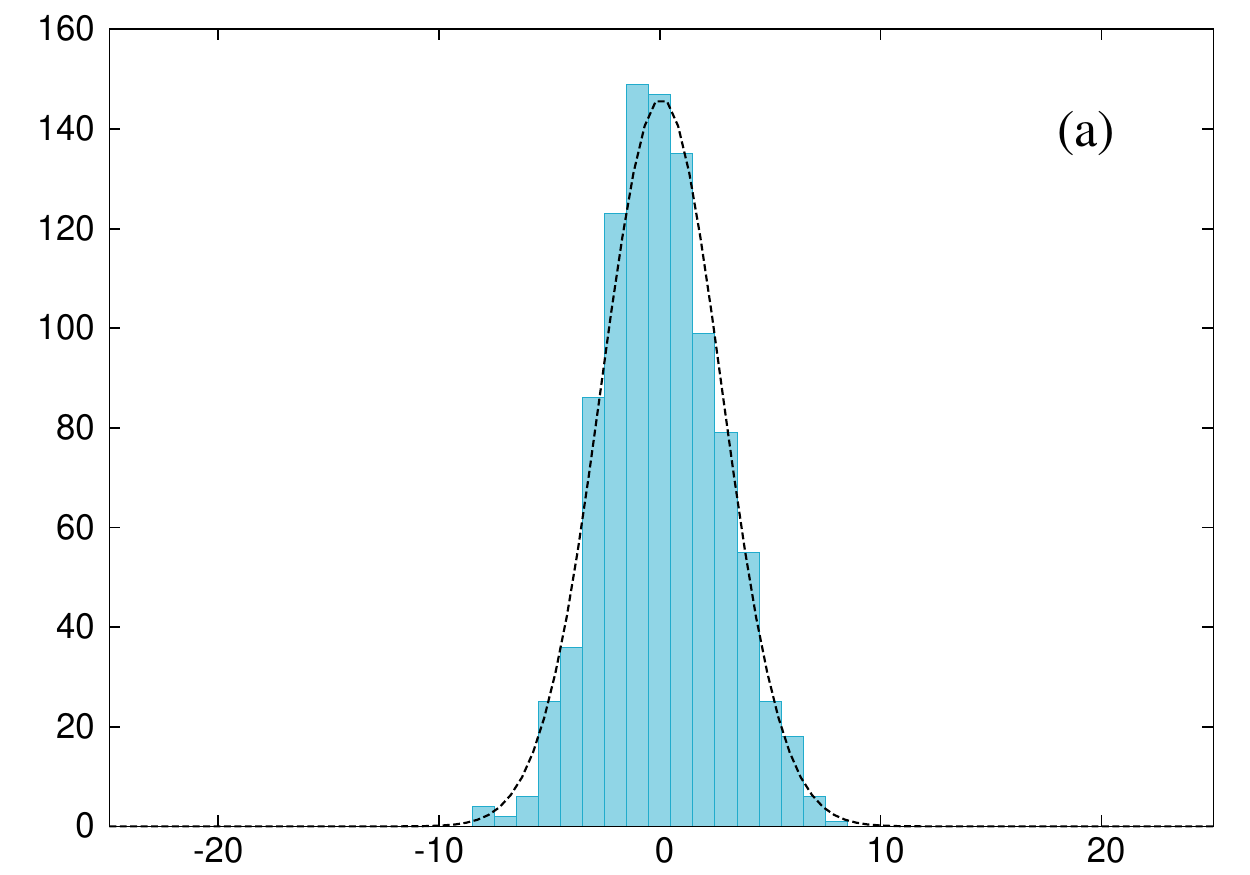}
}
\subfigure{
\includegraphics[width=2.5in]{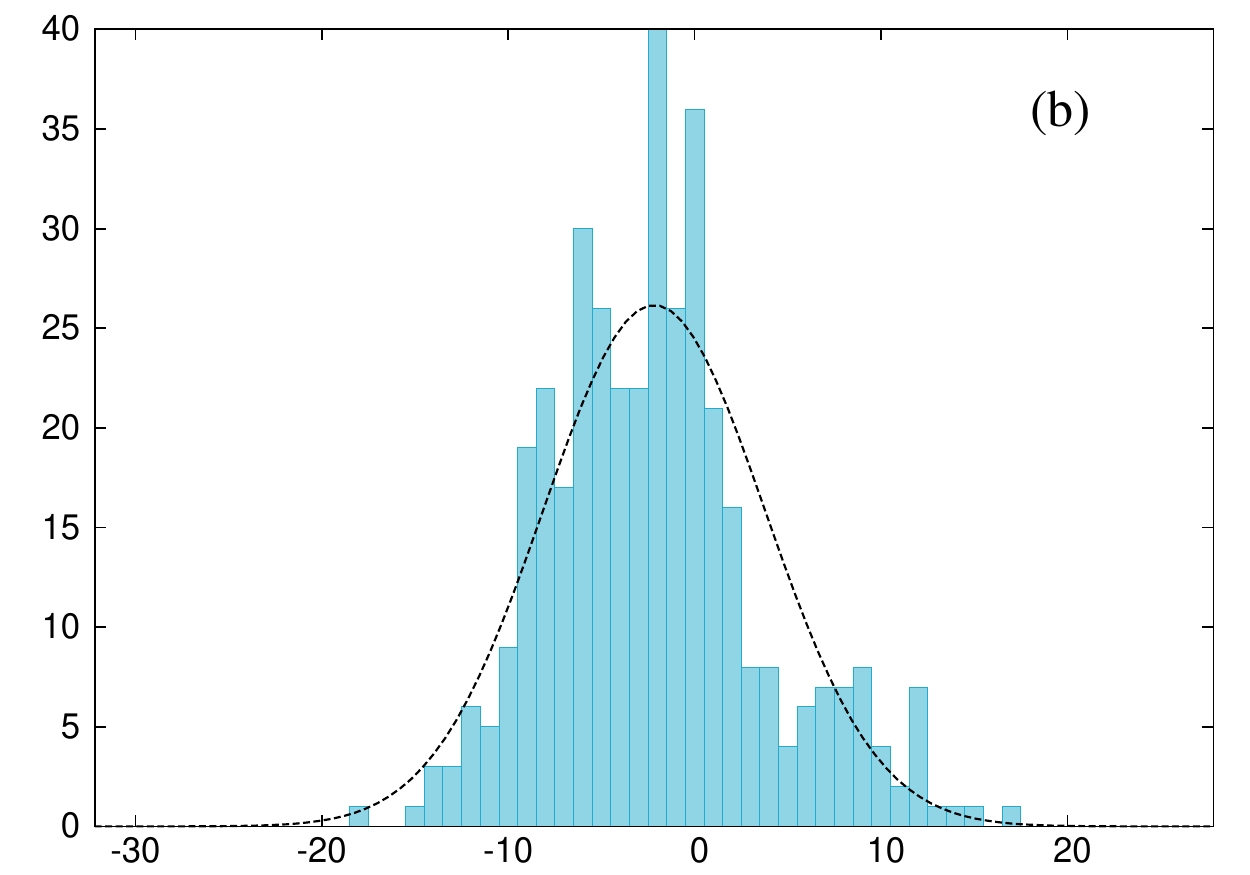}
}

\subfigure{
\includegraphics[width=2.5in]{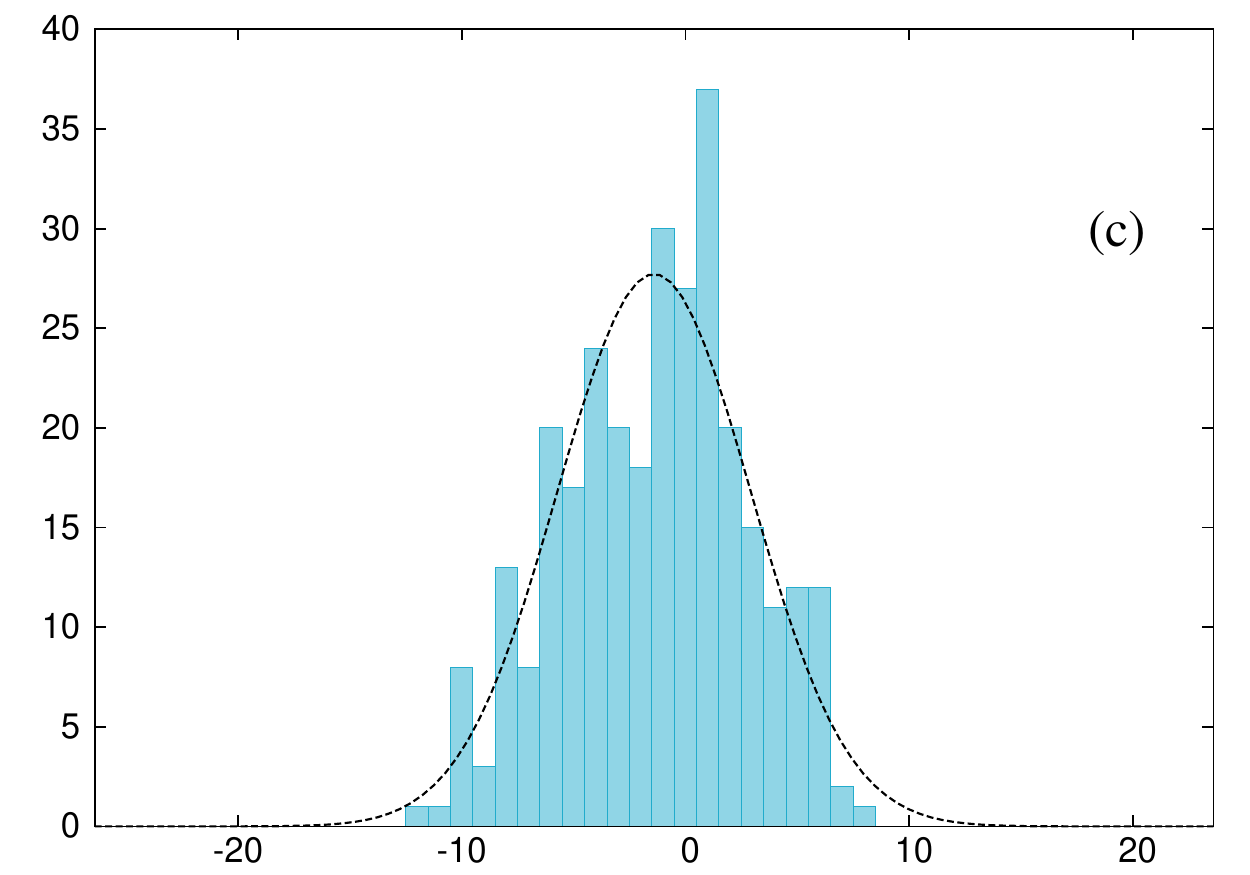}
}
\subfigure{
\includegraphics[width=2.5in]{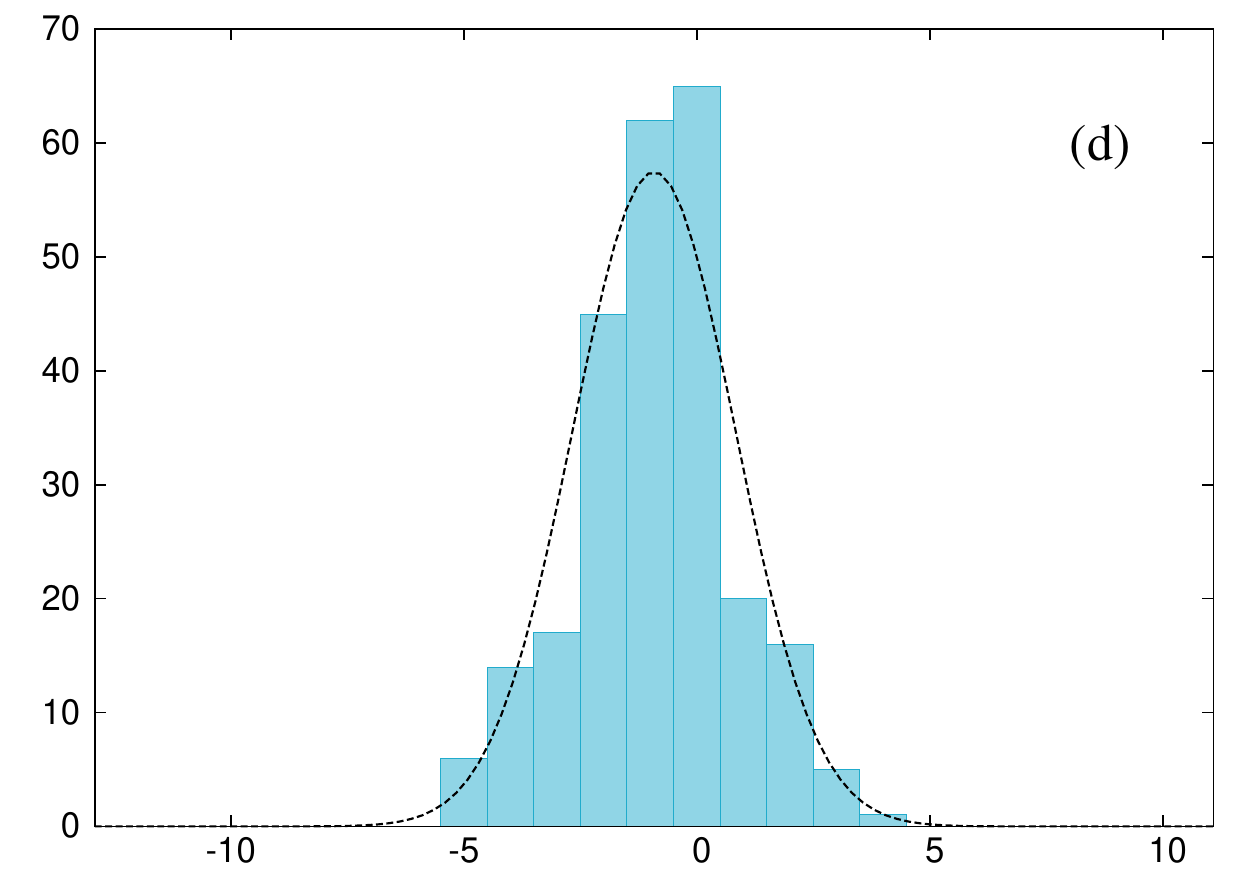}
}
\caption{The topological charge distribution for 
(a) $\beta = 5.6$, $\kappa = 0.1575$, volume $=16^3\times 32$
(b) $\beta = 5.6$, $\kappa = 0.1575$, volume $=24^3\times 48$
(c) $\beta = 5.8$, $\kappa = 0.1543$, volume $=32^3\times 64$
(d) $\beta = 5.8$, $\kappa = 0.15455$, volume $=32^3\times 64$.}
\label{fig5}
\end{figure}

Fig. \ref{fig5} displays four histograms of topological charge distributions,
for two values of $\beta$ and different volumes. 
The topological charge data were put in several bins and 
the bin widths were chosen to be 
unity centered around the integer values of the topological charges for all 
the cases. From theoretical considerations the distribution of the topological 
charge is expected to be a Gaussian \cite{gaussian}. Since 
our configurations are reasonably large in number but finite, an incomplete 
spanning of the topological sectors may occur and    
$\langle Q \rangle$ may  not be zero. Hence we define the susceptibility 
to be 
$
\chi = \frac{1}{V}\left(\langle Q^2\rangle-\langle Q \rangle^2\right).
$

\section{Results}

\begin{figure}
\subfigure
{
\includegraphics[width=3in]{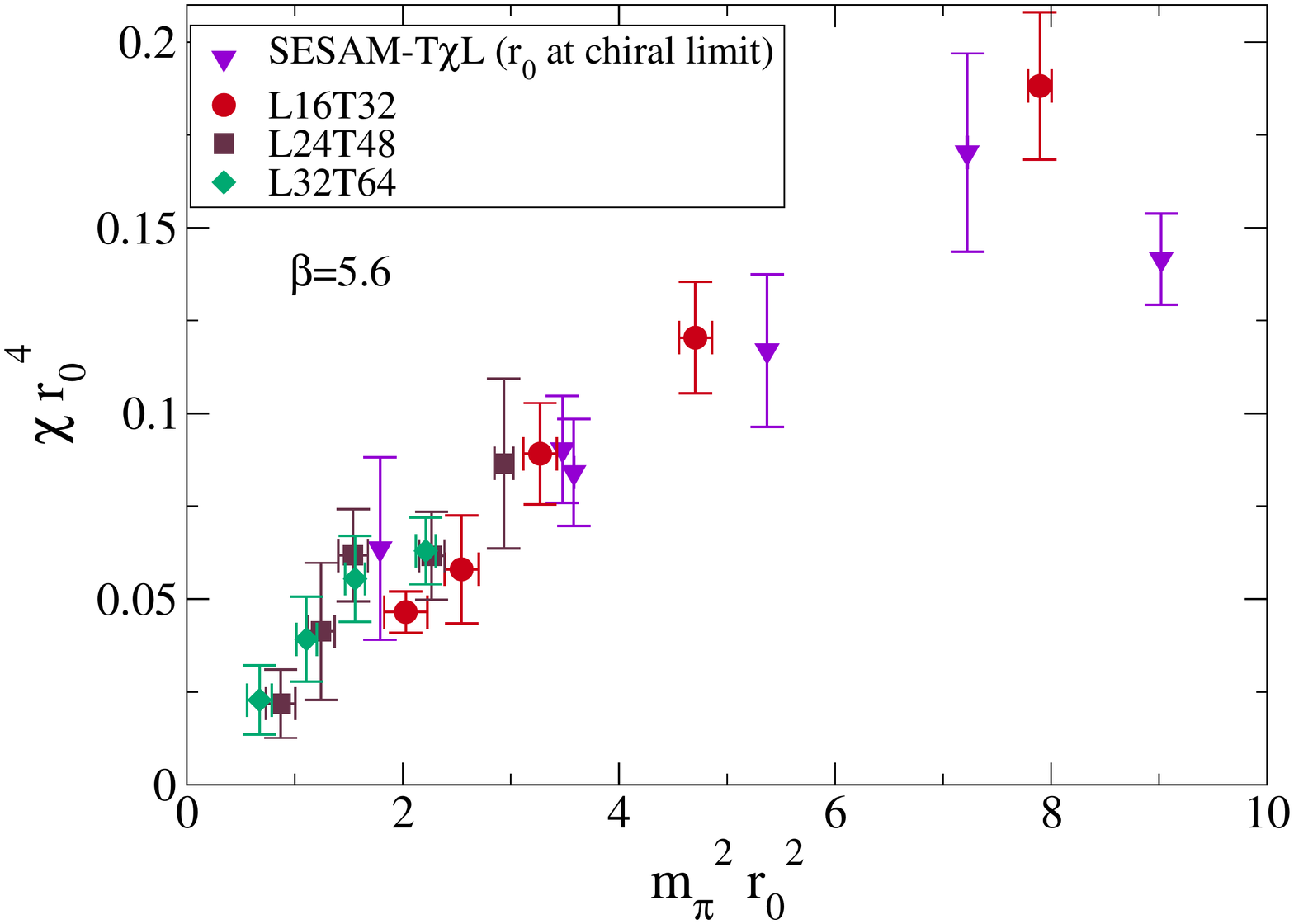}
}
\subfigure
{
\includegraphics[width=3in]{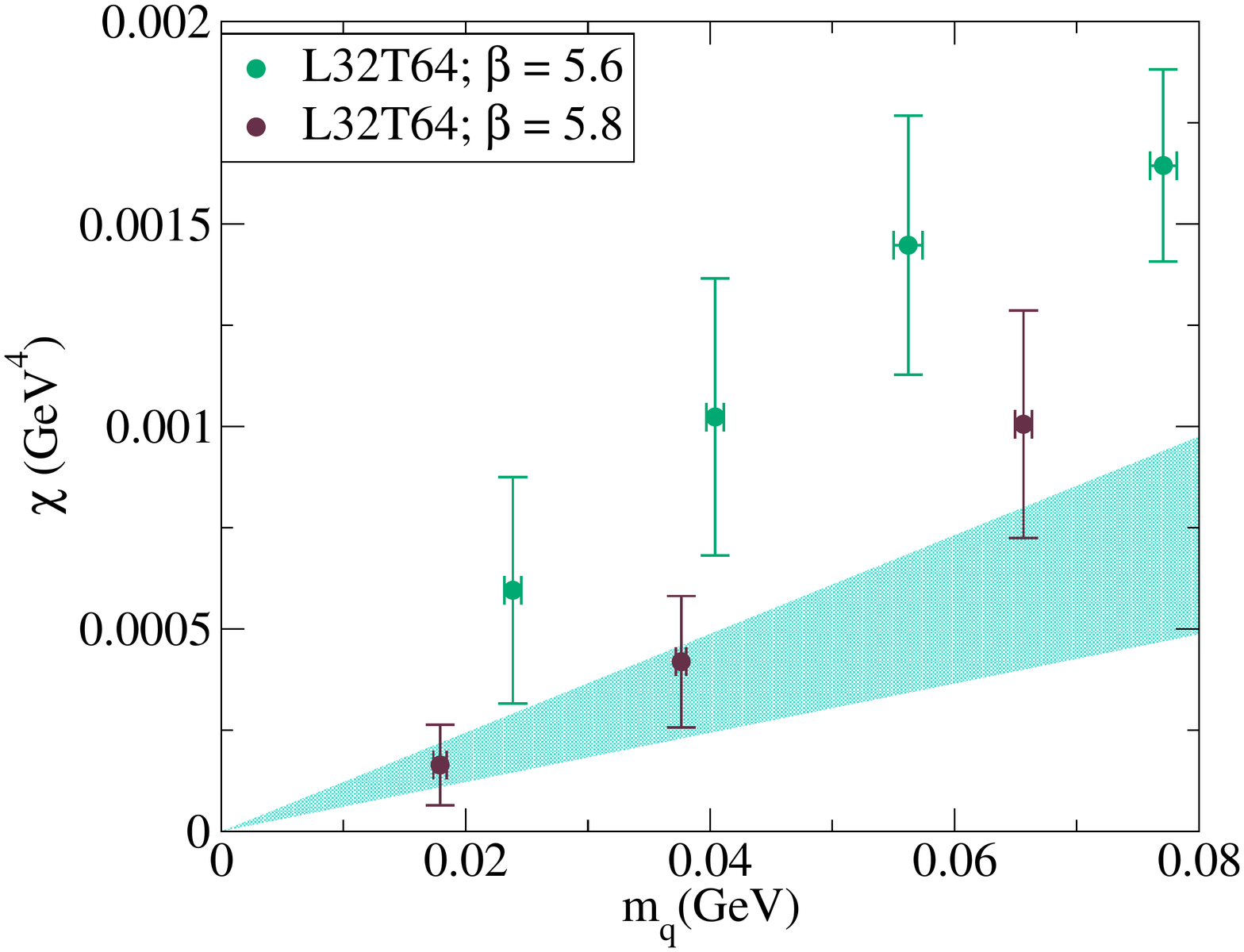}
}
\caption{{\bf Left Figure:} Topological susceptibility  
versus $m_{\pi}^2$ in the units of $r_0$ (at chiral limit) for $\beta = 5.6$ 
and at lattice volumes $16^3 \times 32$, $24^3 \times 48$, and $32^3 \times 64$
compared with the results of SESAM-T$\chi$L collaborations \cite{bali}. 
{\bf Right Figure:} Topological susceptibility versus $m_q$ in the physical units for
$\beta =5.6$ and $\beta = 5.8$ at lattice volume $32 \times 64$. The leading 
order chiral perturbation theory prediction, $\chi=\frac{1}{2}\Sigma~ m_q$
where $\Sigma$ is the chiral condensate, is also shown for the range
$230~\rm{MeV}~\leq\Sigma^{\frac{1}{3}}\leq 290~\rm{MeV}$.    }
\label{fig6}
\end{figure}

 Fig. \ref{fig6} (left) shows our results for topological susceptibility versus 
$m_{\pi}^2$, in the units of Sommer parameter ($r_0$) at the 
chiral limit, for $\beta = 5.6$ and
 at lattice volumes $16^3 \times 32$, $24^3 \times 48$, and $32^3 \times 64$
compared with the results of SESAM-T$\chi$L collaborations \cite{bali}.
We note that SESAM-T$\chi$L results were presented in \cite{bali} by scaling topological susceptibility and $m_\pi^2$ by appropriate powers of quark mass dependent $r_0/a$. Since $r_0/a$ significantly increases with decreasing quark mass, the suppression of topological susceptibility is concealed in such a plot. Using the numbers given in \cite{bali}, we have replotted it after scaling by the value of $r_0$ quoted at the physical point. 
The Fig. \ref{fig6} (left) clearly shows the 
suppression of susceptibility with decreasing quark mass in  the earlier
SESAM-T$\chi$L data with unimproved Wilson fermion. Our results carried out
at larger volume and smaller quark masses unambigously establish the 
suppression of topological susceptibility with decreasing 
quark mass in accordance with
the chiral Ward identity and chiral perturbation theory. 
In  Fig. \ref{fig6} (right) we show  topological susceptibility versus 
nonperturbatively renormalized \cite{becirevic}
quark mass ($m_q$) in $\overline{\rm MS}$ scheme \cite{gimenez} 
at $2$ GeV
in physical units for
$\beta =5.6$ and $\beta = 5.8$ at lattice volume $32 \times 64$. The leading 
order chiral perturbation theory prediction, $\chi=\frac{1}{2}\Sigma~ m_q$
where $\Sigma$ is the chiral condensate, is also shown for the range
$230~\rm{MeV}~\leq\Sigma^{\frac{1}{3}}\leq 290~\rm{MeV}$.

In summary, we have addressed a long standing problem regarding 
topology in lattice simulations
of QCD with unimproved Wilson fermions. Calculations are presented for two 
degenerate flavours.
We have presented integrated autocorrelation time for
both topological charge and topological susceptibility for the two
$\beta$ values studied.
The effects of quark 
mass, lattice volume and the lattice spacing on the spanning of different 
topological sectors are presented.
The suppression of the topological 
susceptibility with respect to  decreasing quark mass, 
expected  from chiral Ward 
identity and chiral perturbation theory is observed.

\vskip .1in
{\bf Acknowledgements}
\vskip .1in
  Numerical calculations are carried out on Cray XD1 and Cray XT5 systems 
supported 
by the 10th and 11th Five Year Plan Projects of the Theory Division, SINP under
the DAE, Govt. of India. We thank Richard Chang for the prompt maintainance of 
the systems and the help in data management. This work was in part based on 
the public lattice gauge theory codes of the 
MILC collaboration \cite{milc} and  Martin L\"{u}scher \cite{ddhmc}.

\end{document}